\documentclass[a4paper,longbibliography,10pt]{article}
\usepackage[T1]{fontenc} 
\usepackage{amsmath}
\usepackage{amssymb}
\usepackage{authblk}
\usepackage[sort&compress,numbers]{natbib}
\usepackage[colorlinks,citecolor=blue,urlcolor=blue,linkcolor=blue]{hyperref}
\newcommand\elm[1]{{\mathop{#1}\limits^{\tiny\textbf{em}}}}

\title{Rotating charged AdS solutions in quadratic $f(T)$ gravity}

\author[a,b]{A. M. Awad \thanks{awad.adel@aucegypt.edu}}
\author[c,d]{G. G. L. Nashed,\thanks{nashed@bue.edu.eg}}
\author[c,d]{W. El Hanafy \thanks{waleed.elhanafy@bue.edu.eg}}

\affil[a]{\small \it Department of Physics, School of Sciences and Engineering, American University in Cairo, P.O. Box 74, AUC Avenue New Cairo, Cairo, Egypt}
\affil[b]{\small \it Department of Physics, Faculty of Science, Ain Shams University, Cairo 11566, Egypt}
\affil[c]{\small \it Centre for Theoretical Physics, the British University in Egypt, El Sherouk City 11837, Egypt}
\affil[d]{\small \it Egyptian Relativity Group, Cairo University, Giza 12613, Egypt}

\date{}

\begin{document}
\maketitle
\begin{abstract}
We present a class of asymptotically anti-de Sitter charged rotating
black hole solutions in $f(T)$ gravity in $N$-dimensions, where
$f(T)=T+\alpha T^{2}$. These solutions are nontrivial extensions of
the solutions presented in \cite{Lemos:1994xp} and
\cite{Awad:2002cz} in the context of general relativity. They are
characterized by cylindrical, toroidal or flat horizons, depending
on global identifications. The static charged black hole
configurations obtained in \cite{Awad:2017tyz} are recovered as
special cases when the rotation parameters vanish. Similar to \cite{Awad:2017tyz} the
static black holes solutions have two
different electric multipole terms in the potential with related
moments. Furthermore, these solutions have milder singularities
compared to their general relativity counterparts. Using the
conserved charges expressions obtained in \cite{Ulhoa:2013gca} and
\cite{Maluf:2008ug} we calculate the total mass/energy and the
angular momentum of these solutions.
\end{abstract}
\section{Introduction}\label{S1}
In the last two decades there has been a growing interest in
gravitational solutions with cosmological constant in general
relativity (GR) and its extensions. This interest has been generated
by seminal observational and theoretical breakthroughs, namely, the
discovery of cosmic acceleration
\cite{Riess:1998cb,Perlmutter:1998np} and the gauge/gravity
dualities \cite{Maldacena:1997re}. Black hole solutions play a very
important role in unraveling several classical and quantum
mechanical aspects of the underlying gravitational theory.
Therefore, it is viewed as an important tool to study various
extensions of GR. Contrary to asymptotically flat black holes,
asymptotically de Sitter (dS) and anti-de Sitter (AdS) black hole
solutions possess more than one type of horizon topology. They could
have spherical, hyperbolic, or flat horizons. dS and AdS black hole
solutions have been obtained and studied in GR extensively, as well
as teleparallel gravities, please see
\cite{Lemos:1994xp,Hawking:1998kw,Chamblin:1999hg,Awad:2002cz,Nashed:2003ee,Hanafy:2015yya,Klemm:1997ea,Iorio:2012cm,2012ChPhL..29e0402G,Xie:2013vua,
Awad:2005ff,Awad:1999xx,Awad:2000ac}, for diverse black hole
solutions.

Since the confirmation of the above cosmological observations there
have been several proposed extensions of GR which are based on
Riemannian as well as other types of geometries. Gravitational
theories based on Riemannian geometry have been extended through
$f(R)$ gravitational theory which was proposed in
\cite{Nojiri:2007as,Bamba:2008ut}. In such a theory, the Ricci
scalar $R$ is replaced by an arbitrary function $f(R)$ in
Einstein-Hilbert action. Other extensions consider a Lagrangian
density on the form of $f(R,{\cal T})$ where ${\cal T}$ the trace of
the energy-momentum tensor of the matter component
\cite{Harko:2011kv}, or some $f(R,G)$ where $G$ is Gauss-Bonnet
scalar
\cite{Cognola:2006eg,Bamba:2011pz,Bamba:2010wb,Bamba:2012vg,Myrzakulov:2010vz}.
Different approach, however, has been developed within
Weitzenb\"{o}ck geometry by introducing the teleparallel torsion
scalar, $T$, as the Lagrangian density instead of the Ricci scalar,
that is the teleparallel equivalent of general relativity (TEGR)
theory. Motivated by the $f(R)$ gravity extension, TEGR has been
generalized to $f(T)$ gravity by replacing $T$ by an arbitrary
function $f(T)$ \cite{DeLaurentis:2015fea}. The $f(T)$ gravity is
considered to be one of the simplest extensions of GR, since its
field equations are still second order
\cite{Bengochea:2008gz,Linder:2010py,Cai:2015emx} in spite of having
arbitrary torsion scalar terms. Although there is an equivalence
between GR and TEGR on the field equations level, their
generalizations $f(R)$ and $f(T)$ are not equivalent.

In general, finding an exact nontrivial black hole solution in the
above extensions, including $f(T)$ gravity, is not an easy task
\cite{Li:2011wu,Li:2010cg,Nashed:2015pda,Nashed:2013bfa,Nashed:2016tbj,
Awad:2017tyz,Capozziello:2012zj,Nashed:uja}.
In this work, we present a rotating black hole in all dimensions
within Maxwell-$f(T)$ theory with a negative cosmological constant,
where $f(T)=T+\alpha T^{2}$. These asymptotically AdS
black holes are characterized by cylindrical, toroidal or flat
horizons depending on the global identifications of some
coordinates. These solutions can be constructed from coordinate
transformations which are allowed locally on a manifold but not
globally \cite{Stachel:1981fg}. They are the $f(T)$ analogue of the
solutions found in GR by Lemos \cite{Lemos:1994xp} and their
generalizations in higher dimensions that were introduced by one of
us in \cite{Awad:2002cz}. The charged static configurations obtained in
\cite{Awad:2017tyz} are recovered in the limit of vanishing rotation
parameters. These interesting black hole solutions have two
different electric multipole terms in the electric potential with
related multipole moments. In addition, they have milder singularities
at $r=0$, similar to that of the static solutions obtained in
\cite{Awad:2017tyz}, compared to Reissner Nordstr\"{o}m solutions in GR. We calculate the energy and the angular
momentum of the black hole using the conserved quantities in the framework of teleparallel gravity.

This work is arranged as follow: In Section \ref{S2}, a brief account of $f(T)$ gravitational theories are provided in addition to the previous solutions
derived in \cite{Awad:2017tyz} within the framework of $f(T)$ gravitational theory. In  Section \ref{S3}, charged rotating N dimensional exact solutions are derived. These solutions have monopoles and quadrupole moments which are not independent, in addition of being asymptotically AdS. In Section \ref{S4}, we calculate the energy and angular momentum of these solutions. In the final section we comment on some physical aspects of these black hole solutions.
\section{Maxwell-$f(T)$ Gravity}\label{S2}
\subsection{Teleparallel geometry}
A Vielbein space can be defined as a pair ($M$, $e_a$), where $M$ is
an $N$-dimensional differentiable manifold and the set $\{e_a\}$
contains $N$ independent vector fields defined globally on $M$, this
set at point $p$ is the basis of its tangent space $T_{p} M$.
Because of the independence of $e_{a}$, the determinant $e\equiv
\det (e_{a}{^{\mu}})$ is nonzero. The vielbein vector fields satisfy
$e_{a}{^{\mu}}e^{a}{_{\nu}}=\delta^{\mu}_{\nu}\quad
\textmd{and}\quad e_{a}{^{\mu}}e^{b}{_{\mu}}=\delta^{b}_{a}$, where
$\delta$ is the Kronecker tensor. Thus, we can construct an
associated (pseudo-Riemannian) metric and its inverse, respectively,
for any set of basis $g_{\mu \nu} \equiv
\eta_{ab}e^{a}{_{\mu}}e^{b}{_{\nu}},\quad g^{\mu \nu} =
\eta^{ab}e_{a}{^{\mu}}e_{b}{^{\nu}}$, where $\eta_{i j}=(-,+,+,+,
\cdots)$ is the metric of $N$-dimensions Minkowski spacetime. Also,
it can be shown that $e=\sqrt{-g}$, where $g\equiv \det(g)$. Thus,
we go further to define the symmetric Levi-Civita connection. In
this sense, the vielbein space is a pseudo-Riemannian as well.
However, if we decide not to use curvature as the basic description
of gravity, we may begin with the vielbein vector fields as the
fundamental field variables. Then, we define the nonsymmetric linear
(Weitzenb\"{o}ck) connection \cite{Wr} $W^{\alpha}{_{\mu\nu}}\equiv
e_{a}{^{\alpha}}\partial_{\nu}e^{a}{_{\mu}}=-e^{a}{_{\mu}}\partial_{\nu}e_{a}{^{\alpha}}$.
This connection is characterized by the property that
$\nabla_{\nu}e_{a}{^{\mu}}\equiv\partial_{\nu}{e_a}^\mu+{W^\mu}_{\lambda
\nu} {e_a}^\lambda\equiv 0$, where the covariant derivative
$\nabla_{\nu}$ is associated to the Weitzenb\"{o}ck connection. This
nonsymmetric connection uniquely determines the teleparallel
geometry, since the vielbein vector fields are parallel with respect
to it. Indeed, the Weitzenb\"{o}ck connection is curvature free, but
it has a non vanishing torsion ${T^\alpha}_{\mu \nu} =
{W^\alpha}_{\nu \mu}-{W^\alpha}_{\mu \nu} ={e_i}^\alpha
[\partial_\mu{e^i}_\nu-\partial_\nu{e^i}_\mu]$. Now we can go
directly to construct the teleparallel torsion scalar
\begin{equation}\label{tor}
T ={T^\alpha}_{\mu \nu} {S_\alpha}^{\mu \nu},\end{equation}
where the superpotential
tensor is defined as ${S_\alpha}^{\mu \nu} :=
\frac{1}{2}\left({K^{\mu\nu}}_\alpha+\delta^\mu_\alpha{T^{\beta
\nu}}_\beta-\delta^\nu_\alpha{T^{\beta \mu}}_\beta\right)$ and the
Contortion tensor is $K_{\alpha \mu
\nu}=\frac{1}{2}\left(T_{\nu\alpha\mu}+T_{\alpha\mu\nu}-T_{\mu\alpha\nu}\right)$.

\subsection{The theory}
We take the action of the $f(T)$-Maxwell theory in $N$-dimensional for asymptotically (Anti)-de-Sitter spacetimes as
\begin{equation}\label{q7}
\mathcal{S}_{g}+\mathcal{S}_{em}=\frac{1}{2\kappa}\int d^{N}x~ |e|\left(f(T)-2\Lambda\right)-\frac{1}{2\kappa}\int d^{N}x~|e| { F}\wedge ^{\star}{F},
\end{equation}
where $\Lambda=-\frac{(N-1)(N-2)}{2l^2}$ is the $N$-dimensional
cosmological constant in $N$ dimensions, $l$ is the length scale of
AdS spacetime, $\kappa$  is a dimensional constant which can be
related to the Newton constant $G_N$ by $\kappa
=2(N-3)\Omega_{N-2} G_N$, where $\Omega_{N-2} =
\frac{2\pi^{(N-1)/2}}{\Gamma([N-1]/2)}$ is the volume of
$(N-2)$-dimensional unit sphere and $\Gamma$ function being the
argument that depends on the dimension of the
spacetime\footnote{For $N= 4$, one can recover $2(N-3)\Omega_{N-2}
= 8 \pi G_4$.}$^{,}$\footnote{The spacetime indices are given by $\mu,\,
\nu \cdots$ and the SO(3,1) indices are given by $a,\, b,\, \cdots$
in which all of them run from 0  \hspace*{0.41cm} to 3. The Latin indices $i, j,
\cdots$ are denote to the SO(3,1) spatial components.}. Also, in the
Maxwell action, $F = d{\cal A}$, with ${\cal A}={\cal A}_{\mu}dx^\mu$ being the
gauge potential 1-form \cite{Awad:2017tyz,Capozziello:2012zj}.

Varying the action (\ref{q7}) with respect to the vielbein and the vector potential ${\cal A}_\mu$, one gets, respectively, the field equations \cite{Bengochea:2008gz}
\begin{eqnarray}\label{q8}
& & \mathfrak{I{^\nu}{_\mu}}={S_\mu}^{\rho \nu} \partial_{\rho} T
f_{TT}+\left[e^{-1}{e^a}_\mu\partial_\rho\left(e{e_a}^\alpha
{S_\alpha}^{\rho \nu}\right)-{T^\alpha}_{\lambda \mu}{S_\alpha}^{\nu \lambda}\right]f_T-\frac{\delta^\nu_\mu}{4}\left(f+\frac{(N-1)(N-2)}{l^2}\right) +{\kappa \over 2} \elm{\mathfrak{T}}{^\nu}{_\mu},\nonumber\\
&&\partial_\nu \left( \sqrt{-g} F^{\mu \nu} \right)=0,
\end{eqnarray}
where $f := f(T)$, $f_{T}:=\frac{\partial f(T)}{\partial T}$, $f_{TT}:=\frac{\partial^2 f(T)}{\partial T^2}$ and $\elm{\mathfrak{T}}{^\nu}{_\mu}$ is the energy
momentum tensor of the electromagnetic field which is given by \cite{Capozziello:2012zj}
\[
\elm{\mathfrak{T}}{^\nu}{_\mu}=F_{\mu \alpha}F^{\nu \alpha}-\frac{1}{4} \delta_\mu{}^\nu F_{\alpha \beta}F^{\alpha \beta}.\]

\subsection{AdS charged black holes with flat horizons}
In a previous work \cite{Awad:2017tyz} we have introduced the
following diagonal vielbein which describes a static configuration
in $N$-dimensions with the coordinates ($t$, $r$, $\phi_1$,
$\phi_2$, $\cdots$, $\phi_{n}$, $z_1$, $z_2$ $\cdots$ $z_k$, $k=1,2
\cdots$ $N-n-2$)
\begin{equation}\label{tetrad}
\hspace{-0.3cm}\begin{tabular}{l}
  $\left({e^{i}}_{\mu}\right)=\left( \sqrt{A(r)}, \; \frac{1}{\sqrt{B(r)}}, \; r, \; r, \; r\;\cdots \right)$,
\end{tabular}
\end{equation}
where $0\leq r< \infty$, $-\infty < t < \infty$, $0\leq \phi_{n}< 2\pi$ and $-\infty < z_k < \infty$. The functions $A(r)$ and $B(r)$ are two unknown functions of the radial coordinate $r$. Thus, the spacetime which can be generated by (\ref{tetrad}) is
\begin{equation}\label{m2}
ds^2=
-A(r)dt^2+\frac{1}{B(r)}dr^2+r^2\left(\sum_{i=1}^{n}d\phi^2_i+\sum_{k=1}^{N-n-2}{dz^2_k
\over l^2}\right).
\end{equation}
  Substituting from Eq. (\ref{tetrad}) into Eq. (\ref{tor}), we evaluate the torsion scalar as\footnote{For abbreviation we will write $A(r)\equiv A$,  \ \ $B(r)\equiv B$, \ \ $A'\equiv\frac{dA}{dr}$,$A''\equiv\frac{d^2A}{dr^2}$,$B''\equiv\frac{d^2B}{dr^2}$ and $B'\equiv\frac{dB}{dr}$ .}
\begin{equation}\label{tor1}
{\mathbf T=2(N-2)\frac{A'B}{rA}+(N-2)(N-3)\frac{B}{r^2}}.
\end{equation}
Using the {\it N}-dimensional spacetime of Eq. (\ref{tetrad}) with Eq. (\ref{tor1}) and the vector potential ${\cal A} = \Phi(r) dt$, we obtain the following equations (\ref{q8}):
\begin{eqnarray}
& &{\mathbf \mathfrak{I}^t{}_t= \frac{2(N-2)Bf_{TT} T'}{r}+\frac{(N-2)f_T[2(N-3)AB+rBA'+rAB']}{r^2A}-f+2\Lambda+\frac{2\Phi'^2(r)B}{A}=0},\nonumber\\
& & {\mathbf \mathfrak{I}^r{}_r= 2Tf_T+2\Lambda-f+\frac{2\Phi'^2(r)B}{A}=0},\nonumber\\
& & {\mathbf\mathfrak{I}^{\phi_1}{}_{\phi_1}= \mathfrak{I}^{\phi_2}{}_{\phi_2}=\cdots \mathfrak{I}^{\phi_n}{}_{\phi_n}=\mathfrak{I}^{z_1}{}_{z_1}=\mathfrak{I}^{z_2}{}_{z_2}\cdots =\mathfrak{I}^{z_k}{}_{z_k}=  \frac{f_{TT} [r^2T+(N-2)(N-3)B]T'}{(N-2)r}+\frac{f_T}{2r^2{A}^2}\Biggl\{2r^2ABA''}\nonumber\\
& & {\mathbf-r^2BA'^2+4(N-3)^2A^2B+2(2N-5)rABA'+r^2AA'B'+2(N-3)rA^2B'\Biggr\}-f+2\Lambda-\frac{2\Phi'^2(r)B}{A}=0}, \nonumber\\
\end{eqnarray}
where $\Phi'=\frac{d\Phi}{dr}$. The general  {\it N}-dimensional solutions with flat horizons of the Maxwell-$f(T)$
theory, where $f(T)=T+\alpha T^2$ of the above differential equations takes the form \cite{Awad:2017tyz}
\begin{eqnarray}\label{sol}
A(r)&=&r^2\Lambda_{eff}-\frac{m}{r^{N-3}}+\frac{3(N-3)q{}^2}{(N-2)r^{2(N-3)}}+\frac{2\sqrt{6\left|\alpha\right|}(N-3)^3q{}^3}{(2N-5)(N-2)r^{3N-8}},
\nonumber\\
B(r)&=&A(r)\left[1+\frac{(N-3)q\sqrt{6\left|\alpha\right|}}{r^{N-2}}\right]^{-2},\label{sol1}\nonumber\\
\Phi(r)&=&\frac{q}{r^{N-3}}+\frac{(N-3)^2q{}^2\sqrt{6\left|\alpha\right|}}{(2N-5)r^{2N-5}}, \end{eqnarray}
where $\Lambda_{eff}= \frac{1}{6(N-1)(N-2)\left|\alpha\right|}$,  $m$ is the mass parameter, $q$ is the charge parameter and $\Phi(r)$ is the electric potential which defines the vector potential ${\cal A}=\Phi(r)dt$. As it is clear from Eq. (\ref{sol}), that the potential $\Phi(r)$ depends on a monopole and quadrupole moments. By setting $q=0$ both momenta vanish and we get a non-charged solution. It is worth mentioning that the solution (\ref{sol}) has been derived for the quadratic polynomial $f(T)$ theory in the presence of the constraint $\Lambda=\frac{1}{24\alpha}$. Consequently, one expects the model parameter to be $\alpha<0$, since the cosmological constant is negative.  The reason of the constraint   $\Lambda=\frac{1}{24\alpha}$ is as follows: If one chooses an ansatz for the charged  solution in which the functions $A(r)$ and $B(r)$ are equal then one gets constant potential, i.e., a trivial potential for a charged solution! In order to avoid this trivial potential, we choose $A(r)= C(r) B(r)$.
In this case the potential will  not be  trivial, but $A(r)$ and $B(r)$ are neither unique nor in closed form. For example, for the 5-dimensional uncharged solution, $\Phi(r)=0$, we have $C(r)=const.$ and $$A(r)=\frac{[1 \pm \sqrt{1-24 \alpha \Lambda}]r^2}{ 72 \alpha}  + \frac{c_1}{r^2}$$  which shows that $A(r)$ is not unique. An extra complication is obtained when the potential is not constant, in this case, $A(r)$ and $B(r)$ can not be expressed in a closed form. Choosing  $\Lambda=\frac{1}{24\alpha}$ leave the solution unique and in a closed form.

Before closing this section, we note that the black hole solution at hand cannot be considered as a special case of the cubic polynomial $f(T)$ gravity which has been studied in \cite{Nashed:2018cth}. In the later, the solution has been obtained under a specific constraint whereas the coefficient of cubic term cannot be made to vanish.
\section{A\lowercase{d}S charged rotating black holes with flat horizons}\label{S3}
One way to add an angular momentum for the above solution in four
dimensions\footnote{It is well known, even in GR, that the addition
of cosmological constant might produce different types of rotating
black holes among them is the class under consideration here, please
see \cite{Klemm:1997ea} for a discussion on these types of rotating
black holes}
\begin{equation}\label{m2}
ds^2= -A(r)dt^2+\frac{1}{B(r)}dr^2+r^2\left(d\phi^2+{dz^2 \over
l^2}\right).
\end{equation}
We follow the procedure developed in
\cite{Lemos:1994xp,Awad:2002cz}, applying the transformations
\begin{equation}\label{t1}
\bar{\phi} =-\Xi~ {\phi}+\frac{ \omega}{l^2}~t,\qquad \qquad \qquad
\bar{t}= \Xi~ t-\omega~ \phi.
\end{equation}
We note that these transformations are allowed locally but not globally on a manifold as will be clarified below. Thus the spacetime (\ref{m2}) reads
\begin{equation}\label{m1}
ds^2=-A(r)\left[\Xi d\bar{t}  -  \omega d\bar{\phi}
\right]^2+\frac{dr^2}{B(r)}+\frac{r^2}{l^4} \left[\omega
d\bar{t}-\Xi l^2 d\bar{\phi} \right]^2+ {r^2 \over l^2} dz^2,
\end{equation}
where
\[\Xi:=\sqrt{1+\frac{\omega^2}{l^2}}.\]
According to Stachel \cite{Stachel:1981fg} if the first Betti number
of the manifold is non-vanishing, which is the case for the
equivalent Riemannian manifold of these solutions, there are no
global diffeomorphisms that can map one of these metrics to the
other, leaving the new manifold with an additional parameter
``$\omega$''. Since in $N$ dimensions we have more than one rotation
parameter, the construction of the rotating tetrad or metric is not
as obvious as the one rotation parameter case as was shown in
\cite{Awad:2002cz}. It requires the addition of other terms which
are not obtained by the above coordinate transformations. In the
higher dimensional case the proposed form of the tetrad for more
than one rotation parameter is given by

\begin{eqnarray}\label{tetrad1}
\left({e^{i}}_{\mu}\right)=\left(
  \begin{array}{cccccccccccccc}
    \Xi\sqrt{A(r)} & 0 &  -\omega_1\sqrt{A(r)}&-\omega_2\sqrt{A(r)}\cdots & -\omega_{n}\sqrt{A(r)}&0&0&\cdots&0 \\[5pt]
    0&\frac{1}{\sqrt{B(r)}} &0 &0\cdots &0&0&0&\cdots & 0\\[5pt]
          \frac{\omega_1r}{l^2} &0 &-\Xi r&0  \cdots &0&0&0&\cdots & 0\\[5pt]
        \frac{ \omega_2r}{l^2} &0 &0  &-\Xi r\cdots & 0&0&0&\cdots & 0\\[5pt]
        \vdots & \vdots  &\vdots&\vdots&\vdots &\vdots&\vdots& \cdots & \vdots \\[5pt]
  \frac{ \omega_nr}{l^2}  &  0 &0&0 \cdots & -\Xi r&0&0&\cdots & 0 \\[5pt]
   0 &  0  &0&0 \cdots &0&r&0&\cdots & 0\\[5pt]
     0 &  0  &0&0 \cdots &0&0&r&\cdots & 0\\[5pt]
       0 &  0 &0&0 \cdots &0&0&0&\cdots & r\\
  \end{array}
\right),
\end{eqnarray}
where $n= \lfloor(N - 1)/2\rfloor$ is the number of rotation
parameters with $\lfloor y \rfloor$ is the integer part of $y$,
$\omega_j$ are the rotation parameters and $\Xi$ is defined
as
\[\Xi:=\sqrt{1+\sum\limits_{j=1}^{{n}}\frac{ {\omega_j}^2}{l^2}}.\]
Also, the functions $A(r)$ and $B(r)$ are given by (\ref{sol}). In addition, the gauge potential takes the form
\begin{equation}\label{Rotpot}
\bar{\Phi}(r)=-\Phi(r)\left[\omega_i~ d\bar{\phi}_i-\Xi~ d\bar{t}\,\right] .
\end{equation}
We note that Eqs. (\ref{sol}) and (\ref{Rotpot}) are also solutions of the stationary configuration (\ref{tetrad1}). Since transformation (\ref{t1}) mixes compact and noncompact coordinates, it leaves the local properties of spacetime the same. However, it does change the spacetime properties globally, c.f. \cite{Lemos:1994xp}.
On other words, the vielbein (\ref{tetrad}) and (\ref{tetrad1}) can be locally mapped into each
other but not globally \cite{Lemos:1994xp,Awad:2002cz}.  One can show that the spacetime which is generated by the
vielbein (\ref{tetrad1}) takes the form
\begin{equation}\label{m1}
ds^2=-A(r)\left[\Xi d\bar{t}  -\sum\limits_{i=1}^{n}  \omega_{i}d\bar{\phi} \right]^2+\frac{dr^2}{B(r)}+\frac{r^2}{l^4}\sum\limits_{i=1 }^{n}\left[\omega_{i}d\bar{t}-\Xi l^2 d\bar{\phi}_i\right]^2+ {r^2 \over l^2} d\Sigma^2-\frac{r^2}{l^2}\sum\limits_{i<j }^{n}\left(\omega_{i}d\bar{\phi}_j-\omega_{j}d\bar{\phi}_i\right)^2,
\end{equation}
where $0\leq r< \infty$, $-\infty < t < \infty$, $0 \leq \phi_{i}< 2\pi$, $i=1,2 \cdots n$ and $-\infty < z_k < \infty$, $d\Sigma^2=dz^kdz^k$ is
the Euclidean metric on ($N-n-2$)-dimensions and $k = 1,2,\cdots, N-3$. We note that the static configuration (\ref{m2}) can be recovered as a special case when the rotation parameters $\omega_j$ are chosen to be vanished.
These charged rotating solutions do not correspond to any known solutions in GR or TEGR since by sending $\alpha \rightarrow 0$ we do not get a well defined tetrad or metric.
Notice that upon setting the mass parameter $m=0$ and the charge $q=0$, the line-element (\ref{m1}) reduces to the $N$-dimensional AdS metric in an unusual coordinate system. One can easily check that the resulting boundary metric is indeed Minkowski through checking the vanishing of its torsion components. Furthermore, this shows that the whole metric in this limit (i.e., line-element with $m=0$ and the charge $q=0$) is the AdS metric.
In the next section, we are going to study the main feature of solution (\ref{tetrad1}).
\section{Conserved Charges}\label{S4}
\subsection{Four-momentum}
Before we calculate the energy or total mass of these black holes, let us follow \cite{Ulhoa:2013gca} deriving the conserved four-momentum for $f(T)$ gravity in few lines.
Variation of the action (\ref{q7}) with respect to the vielbein gives the field equations in the form
\begin{equation}\label{q88}
{S_\mu}^{\rho \nu} \partial_{\rho} T
f_{TT}+\left[e^{-1}{e^a}_\mu\partial_\rho\left(e{e_a}^\alpha
{S_\alpha}^{\rho \nu}\right)-{T^\alpha}_{\lambda \mu}{S_\alpha}^{\nu
\lambda}\right]f_T
-\frac{\delta^\nu_\mu}{4}\left(f+\frac{(N-1)(N-2)}{l^2}\right)  =-{\kappa \over 2}{\mathfrak{T}}{^\nu}{_\mu},
\end{equation}
where  ${\mathfrak{T}}{^\nu}{_\mu}$ is the energy-momentum of the matter.
Equation (\ref{q88}) can be rewritten as
\begin{eqnarray}\label{q888}
\partial_\rho\left(eS^{a \nu \rho }f_T \right)={\kappa \over 2} |e| \left(t^{a \nu}+T^{a \nu}\right), \end{eqnarray}
where \begin{equation}t^{a \nu}={2 \over \kappa}\left[ f_TS^{b c \nu } T_{b c}{}^a-\frac{\delta^\nu_\mu}{4}\left(f+\frac{(N-1)(N-2)}{l^2}\right) \right].\end{equation}\\
Taking the derivative of Eq. (\ref{q888}) with respect to $x^\nu$, we
get
\begin{eqnarray}\label{con}
\partial_\nu\partial_\rho\left(eS^{a \nu \rho }f_T \right)=0 \qquad \textrm {which leads to} \qquad \partial_\nu\left[ {\kappa \over 2} |e| \left(t^{a \nu}+T^{a \nu}\right)\right]=0.  \end{eqnarray}
Equations (\ref{con}) give the conserved $N$-momentum of $f(T)$
gravitational theory in the form
\begin{eqnarray}\label{energy}
P^a= \int_{V} d^{N-1}x |e| t^{0 a}.
\end{eqnarray} Equation (\ref{energy}) which defines the $N$-momentum of $f(T)$ gravity was derived before in \cite{Ulhoa:2013gca}. This has been used, mostly, to calculate energy for asymptotically flat spacetime background. However, the solutions (\ref{tetrad1}) are asymptotically AdS. Here we adopt the point of view of the authors in \cite{PhysRevD.15.2752,Gibbons:1976pt,Gibbons:1978ac,Nashed:2011fg} to calculate conserved quantities of a gravitational solution in reference to a specific background spacetime. These backgrounds are naturally chosen as Minkowski spacetime for asymptotically flat solutions and AdS or dS for asymptotically AdS or dS solutions. Furthermore, infinities due to the asymptotic regions are canceled out in this subtraction prescription leaving the physical quantities finite. For example, the total energy of an AdS black hole, measured by a stationary observer at very large radial distance, is considered to be the difference in energy between the AdS black hole and the AdS space itself. Therefore, in calculating the conserved quantities, it is natural to subtract the contribution due to pure AdS spacetime from that of the solution. Therefore Eqs. (\ref{energy}) and (\ref{an}) take the form
\begin{equation}
P^a=\int_{V} d^{N-1}x  \left[|e|t^{0 a}\right]_{reg},
\end{equation}
where the subscript ``\textit{reg}'' stands for the regularized value of the physical quantity.

Let us now calculate the energy related to the rotating charged
black holes given by Eqs. (\ref{tetrad1}). Using Eq. (\ref{energy}),
it is possible to derive  the components  that are necessary for the
calculations of energy  in the form\footnote{The square parentheses
in the quantities  $S^{(0)(0)1}$  refer to the tangent components,
i.e., $S^{(0)(0)1}=e^0{}_\mu e^0{}_\nu S^{\mu \nu 1}$.}:
\begin{equation}\label{se}  S^{(0)(0)1}=\frac{(N-2){B}}{2r}.  \end{equation}
\begin{equation} \label{en1} P^0=E=\frac{(N-2)[m-\Lambda_{eff}\,r^{(N-1)}]\Xi}{3(N-3)G_N}+\left(\frac{1}{r}\right)+..., \end{equation}where $n\geq 1$.
expression of Eq. (\ref{energy}) takes the form of a surface
integral
\begin{equation} \label{enr} P^a_{reg}:=\frac{2}{\kappa}\int_{\partial V} d^{N-2}x  \left[e{S}^{a 0
\mu}\, n_{\mu} f_T\right]-\frac{2}{\kappa}\int_{\partial V} d^{N-2}x
\left[e{S}^{a 0 \mu}\, n_{\mu}\,f_T\right]_{AdS},\end{equation}where
$n_{\mu}$ is the normal vector to the surface $\partial V$ and \textit{AdS} means evaluating the second expression of Eq. (\ref{energy})
for pure Anti-de-Sitter space. Using (\ref{enr}) in solution
(\ref{tetrad1}), we get
\begin{equation} \label{en11}
E_{reg} = \frac{2(N-2)\,\Xi\,M}{3\,(N-3)},
\end{equation}
where the mass parameter is taken to be $m=2\,G_N\,M$. As expected, the black hole energy is fully characterized by its mass.

\subsection{Angular momentum}

Although there is a hamiltonian formulation in teleparallel equivalent of general relativity (i.e., $f(T)=T$) which produces some known expressions for the conserved four-momenta and
angular momentum \cite{Maluf:2008ug}, there is no known expression for angular momentum in $f(T)$ gravity. But since the angular momentum is independent of the charge $q$, by sending $q\rightarrow 0$ we obtain a solution with constant torsion scale $T$, since the scalar torsion is given by,
\begin{equation}
T = {-1 \over 6 \alpha}+{2q\sqrt{6} \over 3 \sqrt{\alpha}\,\, r^3}.
\end{equation}
A solution in $f(T)$ gravity with constant torsion scalars, $T=T_c$, is equivalent to a solution in TEGR with constant torsion, where $T'=f(T_c)$. Therefore, one can use Maluf's expression in \cite{Maluf:2008ug} to calculate the angular momentum of our solution in this limit.
Following \cite{Maluf:2008ug}, the angular momentum tensor can be written in terms of the superpotential $S^{a b c}$ in the following form
\begin{eqnarray}\label{sig}
 M^{a \mu c}\equiv |e| e_b{}^\mu  \left[S^{a b c}-S^{c a b}\right]=|e|  \left[S^{a \mu c}-S^{c a \mu}\right] =-\frac{1}{2}\partial_\nu\{|e| \,[e^{a\nu}e^{c \mu}-e^{a\mu}e^{c \nu}] \, \}.\end{eqnarray}
From equation (\ref{sig}) one can easily show that
\begin{eqnarray}\label{sig1}
\
\partial_\mu M^{a \mu c}=0.\end{eqnarray}
Using Eqs. (\ref{sig}) and (\ref{sig1}) the conserved angular momentum is given by
\begin{eqnarray}\label{an}
L^{a b}=\int_{V} d^{N-1} M^{a 0 b}&=&-\frac{1}{\kappa}\int_{V} d^{N-1}x \partial_\nu\{|e| \, [e^{a\nu}e^{b 0}-e^{a0}e^{b \nu}] \, \} \nonumber \\
&=&-\frac{1}{\kappa}\int_{\partial V} d^{N-2}x\; n_\rho |e| \, [e^{a\rho}e^{b 0}-e^{a0}e^{b \rho}],
\end{eqnarray}
where $n_\rho$ is the outward unit normal vector.

Now we are going to calculate the angular momentum of solution (\ref{tetrad1}) in the limit $q\rightarrow 0$. For this aim we are going to list the necessary components needed for these calculations. The non-vanishing components of the torsion tensor, $T^{a b c}=e^a{}_\mu e^b{}_\nu e^c{}_\rho T^{\mu \nu \rho}$, and the superpotential tensor, $S^{a b c}=e^a{}_\mu e^b{}_\nu e^c{}_\rho   S^{\mu \nu \rho}$, are
\begin{eqnarray} \label{se1}
&&T_{(0) (1)  (0)  }=\frac{A'\sqrt{B}}{2A},\qquad\\[5pt]
&&T_{(N-i) (N-j) (1) }=\frac{\Bigg(\Bigg[l^2\Xi^2-\sum\limits_{a=1}^{n}\omega_a{}^2\Bigg]\delta_{i j}+\omega_i\omega_j\Bigg)\sqrt{B}}{l^2\Xi^2 r},\qquad \\[5pt]
&&T_{(N-n-\sum\limits_{k=1}^{(N-n-2)}\;\;k) (N-n-\sum\limits_{k=1}^{(N-n-2)}\;\;k) (1) }=\frac{\sqrt{B}}{r},\qquad \\[5pt]
&&S_{(0) (0) 1}=\frac{(N-2){B}}{2r}, \\[5pt]
&&S_{(N-i) (1) (N-j)  }=\frac{\sqrt{B}\Bigg(\Bigg[l^2\Xi^2-\sum\limits_{a=1}^{n}\omega_a{}^2\Bigg]\delta_{i j}+\omega_i\omega_j\Bigg)\Bigg[2(N-3)A+rA'\Bigg]}{4Al^2\Xi^2 r},\qquad   \\[5pt]
&&S_{(N-n-\sum\limits_{k=1}^{(N-n-2)}\;\;k) (1) (N-n-\sum\limits_{k=1}^{(N-n-2)}\;\;k)  }=\frac{\sqrt{B}\Bigg[2(N-3)A+rA'\Bigg]}{4A r}.\qquad
\end{eqnarray}
Similar to the energy calculations, we are going to use the background subtraction prescription to calculate the angular momentum of the black hole relative to the AdS space background.

\begin{equation}
L^{i j} =-\frac{1}{\kappa}\int_V d^{N-1}x e^i{}_\mu e^j{}_\nu  |e|
\left[(S^{\mu 0\nu}-S^{\nu 0 \mu})\right]_{reg}.
\end{equation}
Using the above equation one gets
\begin{equation} \label{an1}
{J_i|}_{reg}=\frac{ \omega_i M }{2\, (N-3)},
\end{equation}
where
\begin{equation}\label{an12} {J_i}=\epsilon_{i j k}L^{j k}.
\end{equation}
As clear from the above equations that the angular momentum vanishes when the rotation parameters $\omega_i$ vanish. In conclusion, under the constraint $q\to 0$, equations (\ref{en11}) and (\ref{an1}) show that the black holes are characterized by their masses and angular momenta.
\section{Conclusions}\label{S5}
In this work, we present a new class of charged rotating solutions
in $f(T)$ theories in $N$ dimensions. These solutions are obtained for
$f(T)=T+\alpha T^2$, where the parameter $\alpha <0$. It is worth to mention that these solutions cannot be considered as special cases of the solutions of the cubic polynomial $f(T)$ gravity which have been recently studied in \cite{Nashed:2018cth}. This is because the later are obtained whereas the cubic contribution is parameterized by an extra parameter which cannot be made to vanish. One of the attractive features of the solutions at hand is that their electric
potential has related monopole and quadrupole moments. The relation
between these moments is a result of demanding an asymptotically
AdS solution. It is intriguing to note that all these
black holes have a singularity at $r=0$, which is milder than that
of their corresponding solutions in TEGR or GR. The asymptotic behavior of
the Kretschmann invariant, the Ricci tensor squared, and the Ricci
scalar have the same form of the charged ones presented in
\cite{Awad:2017tyz}, i.e. $K=R_{\mu \nu}R^{\mu \nu} \sim
r^{-2(N-2)}$, $R\sim r^{-(N-2)}$. This is in contrast with their
corresponding known solutions in Einstein-Maxwell theory in both GR
and TEGR. Also it is important to mention that, in spite that the
 charged rotating black hole has different components for $g_{tt}$ and
$g^{rr}$, their Killing and event horizons coincide.

To understand these solutions more, we calculate their total energy
and angular momentum. For this aim we have used the mass/energy
expression in the framework of $f(T)$ obtained by
\cite{Ulhoa:2013gca}. For the angular momentum we have used the
expression in \cite{Maluf:2008ug}. We have used the form of the
energy-momentum tensor to calculate the total energy of the rotating
charged black holes and have shown that the resulting form depends
on the mass of the black hole which is consistent with the derived
form in \cite{Awad:2017tyz}.

For calculating the angular momentum of the solutions one notices
that, although there is a hamiltonian formulation for TEGR, which
produces a known expression for the angular momentum, there is no
known expression for angular momentum in $f(T)$ gravity. We
argue that since the angular momentum in our solution is independent
of the charge $q$, by sending $q\rightarrow 0$ we obtain a solution
with constant torsion scaler $T$, therefore, one can use the
angular momentum expression for TEGR following \cite{Maluf:2008ug}.
As a results we have used the expressions obtained
in \cite{Ulhoa:2013gca} and \cite{Maluf:2008ug} to calculate the mass
and angular momentum of these solutions together with the
subtraction technique used for asymptotically de-Sitter and
Anti-de-Sitter solutions.
One of the interesting features that we would like to check in future works is that if these milder curvature singularities are weak enough to make these singularities ``Tipler weak'' according to Tipler's criteria \cite{Tipler:1977}. If it is weak enough, this might leads to possible extensions of the manifold as was shown in the same theory (i.e., $f(T)=T+\alpha T^2$) for some cosmological solutions in \cite{Awad:2017sau}.
\section*{acknowledgments}
This work is partially supported by the Egyptian Ministry of Scientific Research under project No. 24-2-12.

\begin{thebibliography}{10}

\bibitem{Lemos:1994xp}
J.~P.~S. Lemos, \emph{{Cylindrical black hole in general relativity}},
  \href{https://doi.org/10.1016/0370-2693(95)00533-Q}{\emph{Phys. Lett.}
  {\bfseries B353} (1995) 46--51},
  [\href{https://arxiv.org/abs/gr-qc/9404041}{{\ttfamily gr-qc/9404041}}].

\bibitem{Awad:2002cz}
A.~M. Awad, \emph{{Higher dimensional charged rotating solutions in (A)dS
  space-times}},
  \href{https://doi.org/10.1088/0264-9381/20/13/327}{\emph{Class. Quant. Grav.}
  {\bfseries 20} (2003) 2827--2834},
  [\href{https://arxiv.org/abs/hep-th/0209238}{{\ttfamily hep-th/0209238}}].

\bibitem{Awad:2017tyz}
A.~M. Awad, S.~Capozziello and G.~G.~L. Nashed, \emph{{$D$-dimensional charged
  Anti-de-Sitter black holes in $f(T)$ gravity}},
  \href{https://doi.org/10.1007/JHEP07(2017)136}{\emph{JHEP} {\bfseries 07}
  (2017) 136}, [\href{https://arxiv.org/abs/1706.01773}{{\ttfamily
  1706.01773}}].

\bibitem{Ulhoa:2013gca}
S.~C. Ulhoa and E.~P. Spaniol, \emph{{On the Gravitational Energy-Momentum
  Vector in f(T) Theories}},
  \href{https://doi.org/10.1142/S0218271813500697}{\emph{Int. J. Mod. Phys.}
  {\bfseries D22} (2013) 1350069},
  [\href{https://arxiv.org/abs/1303.3144}{{\ttfamily 1303.3144}}].

\bibitem{Maluf:2008ug}
J.~W. Maluf and S.~C. Ulhoa, \emph{{On the gravitational angular momentum of
  rotating sources}},
  \href{https://doi.org/10.1007/s10714-008-0701-x}{\emph{Gen. Rel. Grav.}
  {\bfseries 41} (2009) 1233--1247},
  [\href{https://arxiv.org/abs/0810.1934}{{\ttfamily 0810.1934}}].

\bibitem{Riess:1998cb}
{\scshape Supernova Search Team} collaboration, A.~G. Riess et~al.,
  \emph{{Observational evidence from supernovae for an accelerating universe
  and a cosmological constant}},
  \href{https://doi.org/10.1086/300499}{\emph{Astron. J.} {\bfseries 116}
  (1998) 1009--1038}, [\href{https://arxiv.org/abs/astro-ph/9805201}{{\ttfamily
  astro-ph/9805201}}].

\bibitem{Perlmutter:1998np}
{\scshape Supernova Cosmology Project} collaboration, S.~Perlmutter et~al.,
  \emph{{Measurements of Omega and Lambda from 42 high redshift supernovae}},
  \href{https://doi.org/10.1086/307221}{\emph{Astrophys. J.} {\bfseries 517}
  (1999) 565--586}, [\href{https://arxiv.org/abs/astro-ph/9812133}{{\ttfamily
  astro-ph/9812133}}].

\bibitem{Maldacena:1997re}
J.~M. Maldacena, \emph{{The Large N limit of superconformal field theories and
  supergravity}}, \href{https://doi.org/10.1023/A:1026654312961,
  10.4310/ATMP.1998.v2.n2.a1}{\emph{Int. J. Theor. Phys.} {\bfseries 38} (1999)
  1113--1133}, [\href{https://arxiv.org/abs/hep-th/9711200}{{\ttfamily
  hep-th/9711200}}].

\bibitem{Hawking:1998kw}
S.~W. Hawking, C.~J. Hunter and M.~Taylor, \emph{{Rotation and the AdS / CFT
  correspondence}},
  \href{https://doi.org/10.1103/PhysRevD.59.064005}{\emph{Phys. Rev.}
  {\bfseries D59} (1999) 064005},
  [\href{https://arxiv.org/abs/hep-th/9811056}{{\ttfamily hep-th/9811056}}].

\bibitem{Chamblin:1999hg}
A.~Chamblin, R.~Emparan, C.~V. Johnson and R.~C. Myers, \emph{{Holography,
  thermodynamics and fluctuations of charged AdS black holes}},
  \href{https://doi.org/10.1103/PhysRevD.60.104026}{\emph{Phys. Rev.}
  {\bfseries D60} (1999) 104026},
  [\href{https://arxiv.org/abs/hep-th/9904197}{{\ttfamily hep-th/9904197}}].

\bibitem{Nashed:2003ee}
G.~G.~L. Nashed, \emph{{Stability of the vacuum nonsingular black hole}},
  \href{https://doi.org/10.1016/S0960-0779(02)00168-6}{\emph{Chaos Solitons
  Fractals} {\bfseries 15} (2003) 841},
  [\href{https://arxiv.org/abs/gr-qc/0301008}{{\ttfamily gr-qc/0301008}}].

\bibitem{Hanafy:2015yya}
W.~El~Hanafy and G.~G.~L. Nashed, \emph{{Exact Teleparallel Gravity of Binary
  Black Holes}},
  \href{https://doi.org/10.1007/s10509-016-2662-y}{\emph{Astrophys. Space Sci.}
  {\bfseries 361} (2016) 68},
  [\href{https://arxiv.org/abs/1507.07377}{{\ttfamily 1507.07377}}].

\bibitem{Klemm:1997ea}
D.~Klemm, V.~Moretti and L.~Vanzo, \emph{{Rotating topological black holes}},
  \href{https://doi.org/10.1103/PhysRevD.60.109902,
  10.1103/PhysRevD.57.6127}{\emph{Phys. Rev.} {\bfseries D57} (1998)
  6127--6137}, [\href{https://arxiv.org/abs/gr-qc/9710123}{{\ttfamily
  gr-qc/9710123}}].

\bibitem{Iorio:2012cm}
L.~Iorio and E.~N. Saridakis, \emph{{Solar system constraints on f(T)
  gravity}}, \href{https://doi.org/10.1111/j.1365-2966.2012.21995.x}{\emph{Mon.
  Not. Roy. Astron. Soc.} {\bfseries 427} (2012) 1555},
  [\href{https://arxiv.org/abs/1203.5781}{{\ttfamily 1203.5781}}].

\bibitem{2012ChPhL..29e0402G}
G.~L.~N. {Gamal}, \emph{{Spherically Symmetric Solutions on a Non-Trivial Frame
  in f(T) Theories of Gravity}},
  \href{https://doi.org/10.1088/0256-307X/29/5/050402}{\emph{Chinese Physics
  Letters} {\bfseries 29} (May, 2012) 050402},
  [\href{https://arxiv.org/abs/1111.0003}{{\ttfamily 1111.0003}}].

\bibitem{Xie:2013vua}
Y.~Xie and X.-M. Deng, \emph{{$f(T)$ gravity: effects on astronomical
  observation and Solar System experiments and upper-bounds}},
  \href{https://doi.org/10.1093/mnras/stt991}{\emph{Mon. Not. Roy. Astron.
  Soc.} {\bfseries 433} (2013) 3584--3589},
  [\href{https://arxiv.org/abs/1312.4103}{{\ttfamily 1312.4103}}].

\bibitem{Awad:2005ff}
A.~M. Awad, \emph{{Higher dimensional Taub-NUTS and Taub-Bolts in
  Einstein-Maxwell gravity}},
  \href{https://doi.org/10.1088/0264-9381/23/9/006}{\emph{Class. Quant. Grav.}
  {\bfseries 23} (2006) 2849--2860},
  [\href{https://arxiv.org/abs/hep-th/0508235}{{\ttfamily hep-th/0508235}}].

\bibitem{Awad:1999xx}
A.~M. Awad and C.~V. Johnson, \emph{{Holographic stress tensors for Kerr - AdS
  black holes}}, \href{https://doi.org/10.1103/PhysRevD.61.084025}{\emph{Phys.
  Rev.} {\bfseries D61} (2000) 084025},
  [\href{https://arxiv.org/abs/hep-th/9910040}{{\ttfamily hep-th/9910040}}].

\bibitem{Awad:2000ac}
A.~M. Awad and C.~V. Johnson, \emph{{Scale versus conformal invariance in the
  AdS / CFT correspondence}},
  \href{https://doi.org/10.1103/PhysRevD.62.125010}{\emph{Phys. Rev.}
  {\bfseries D62} (2000) 125010},
  [\href{https://arxiv.org/abs/hep-th/0006037}{{\ttfamily hep-th/0006037}}].

\bibitem{Nojiri:2007as}
S.~Nojiri and S.~D. Odintsov, \emph{{Unifying inflation with LambdaCDM epoch in
  modified f(R) gravity consistent with Solar System tests}},
  \href{https://doi.org/10.1016/j.physletb.2007.10.027}{\emph{Phys. Lett.}
  {\bfseries B657} (2007) 238--245},
  [\href{https://arxiv.org/abs/0707.1941}{{\ttfamily 0707.1941}}].

\bibitem{Bamba:2008ut}
K.~Bamba, S.~Nojiri and S.~D. Odintsov, \emph{{The Universe future in modified
  gravity theories: Approaching the finite-time future singularity}},
  \href{https://doi.org/10.1088/1475-7516/2008/10/045}{\emph{JCAP} {\bfseries
  0810} (2008) 045}, [\href{https://arxiv.org/abs/0807.2575}{{\ttfamily
  0807.2575}}].

\bibitem{Harko:2011kv}
T.~Harko, F.~S.~N. Lobo, S.~Nojiri and S.~D. Odintsov, \emph{{$f(R,T)$
  gravity}}, \href{https://doi.org/10.1103/PhysRevD.84.024020}{\emph{Phys.
  Rev.} {\bfseries D84} (2011) 024020},
  [\href{https://arxiv.org/abs/1104.2669}{{\ttfamily 1104.2669}}].

\bibitem{Cognola:2006eg}
G.~Cognola, E.~Elizalde, S.~Nojiri, S.~D. Odintsov and S.~Zerbini, \emph{{Dark
  energy in modified Gauss-Bonnet gravity: Late-time acceleration and the
  hierarchy problem}},
  \href{https://doi.org/10.1103/PhysRevD.73.084007}{\emph{Phys. Rev.}
  {\bfseries D73} (2006) 084007},
  [\href{https://arxiv.org/abs/hep-th/0601008}{{\ttfamily hep-th/0601008}}].

\bibitem{Bamba:2011pz}
K.~Bamba and C.-Q. Geng, \emph{{Thermodynamics of cosmological horizons in
  $f(T)$ gravity}},
  \href{https://doi.org/10.1088/1475-7516/2011/11/008}{\emph{JCAP} {\bfseries
  1111} (2011) 008}, [\href{https://arxiv.org/abs/1109.1694}{{\ttfamily
  1109.1694}}].

\bibitem{Bamba:2010wb}
K.~Bamba, C.-Q. Geng, C.-C. Lee and L.-W. Luo, \emph{{Equation of state for
  dark energy in $f(T)$ gravity}},
  \href{https://doi.org/10.1088/1475-7516/2011/01/021}{\emph{JCAP} {\bfseries
  1101} (2011) 021}, [\href{https://arxiv.org/abs/1011.0508}{{\ttfamily
  1011.0508}}].

\bibitem{Bamba:2012vg}
K.~Bamba, R.~Myrzakulov, S.~Nojiri and S.~D. Odintsov, \emph{{Reconstruction of
  $f(T)$ gravity: Rip cosmology, finite-time future singularities and
  thermodynamics}},
  \href{https://doi.org/10.1103/PhysRevD.85.104036}{\emph{Phys. Rev.}
  {\bfseries D85} (2012) 104036},
  [\href{https://arxiv.org/abs/1202.4057}{{\ttfamily 1202.4057}}].

\bibitem{Myrzakulov:2010vz}
R.~Myrzakulov, \emph{{Accelerating universe from F(T) gravity}},
  \href{https://doi.org/10.1140/epjc/s10052-011-1752-9}{\emph{Eur. Phys. J.}
  {\bfseries C71} (2011) 1752},
  [\href{https://arxiv.org/abs/1006.1120}{{\ttfamily 1006.1120}}].

\bibitem{DeLaurentis:2015fea}
M.~De~Laurentis, M.~Paolella and S.~Capozziello, \emph{{Cosmological inflation
  in $F(R,\mathcal{G})$ gravity}},
  \href{https://doi.org/10.1103/PhysRevD.91.083531}{\emph{Phys. Rev.}
  {\bfseries D91} (2015) 083531},
  [\href{https://arxiv.org/abs/1503.04659}{{\ttfamily 1503.04659}}].

\bibitem{Bengochea:2008gz}
G.~R. Bengochea and R.~Ferraro, \emph{{Dark torsion as the cosmic speed-up}},
  \href{https://doi.org/10.1103/PhysRevD.79.124019}{\emph{Phys. Rev.}
  {\bfseries D79} (2009) 124019},
  [\href{https://arxiv.org/abs/0812.1205}{{\ttfamily 0812.1205}}].

\bibitem{Linder:2010py}
E.~V. Linder, \emph{{Einstein's Other Gravity and the Acceleration of the
  Universe}}, \href{https://doi.org/10.1103/PhysRevD.81.127301,
  10.1103/PhysRevD.82.109902}{\emph{Phys. Rev.} {\bfseries D81} (2010) 127301},
  [\href{https://arxiv.org/abs/1005.3039}{{\ttfamily 1005.3039}}].

\bibitem{Cai:2015emx}
Y.-F. Cai, S.~Capozziello, M.~De~Laurentis and E.~N. Saridakis, \emph{{f(T)
  teleparallel gravity and cosmology}},
  \href{https://doi.org/10.1088/0034-4885/79/10/106901}{\emph{Rept. Prog.
  Phys.} {\bfseries 79} (2016) 106901},
  [\href{https://arxiv.org/abs/1511.07586}{{\ttfamily 1511.07586}}].

\bibitem{Li:2011wu}
B.~Li, T.~P. Sotiriou and J.~D. Barrow, \emph{{Large-scale Structure in f(T)
  Gravity}}, \href{https://doi.org/10.1103/PhysRevD.83.104017}{\emph{Phys.
  Rev.} {\bfseries D83} (2011) 104017},
  [\href{https://arxiv.org/abs/1103.2786}{{\ttfamily 1103.2786}}].

\bibitem{Li:2010cg}
B.~Li, T.~P. Sotiriou and J.~D. Barrow, \emph{{$f(T)$ gravity and local Lorentz
  invariance}}, \href{https://doi.org/10.1103/PhysRevD.83.064035}{\emph{Phys.
  Rev.} {\bfseries D83} (2011) 064035},
  [\href{https://arxiv.org/abs/1010.1041}{{\ttfamily 1010.1041}}].

\bibitem{Nashed:2015pda}
G.~L. Nashed, \emph{{FRW in quadratic form of $f(T)$ gravitational theories}},
  \href{https://doi.org/10.1007/s10714-015-1917-1}{\emph{Gen. Rel. Grav.}
  {\bfseries 47} (2015) 75},
  [\href{https://arxiv.org/abs/1506.08695}{{\ttfamily 1506.08695}}].

\bibitem{Nashed:2013bfa}
G.~G.~L. Nashed, \emph{{Spherically symmetric charged-dS solution in $f(T)$
  gravity theories}},
  \href{https://doi.org/10.1103/PhysRevD.88.104034}{\emph{Phys. Rev.}
  {\bfseries D88} (2013) 104034},
  [\href{https://arxiv.org/abs/1311.3131}{{\ttfamily 1311.3131}}].

\bibitem{Nashed:2016tbj}
G.~G.~L. Nashed and W.~El~Hanafy, \emph{{Analytic rotating black hole solutions
  in $N$-dimensional $f(T)$ gravity}},
  \href{https://doi.org/10.1140/epjc/s10052-017-4663-6}{\emph{Eur. Phys. J.}
  (2017) 90}, [\href{https://arxiv.org/abs/1612.05106}{{\ttfamily
  1612.05106}}].

\bibitem{Capozziello:2012zj}
S.~Capozziello, P.~A. Gonzalez, E.~N. Saridakis and Y.~Vasquez, \emph{{Exact
  charged black-hole solutions in D-dimensional $f(T)$ gravity: torsion vs
  curvature analysis}},
  \href{https://doi.org/10.1007/JHEP02(2013)039}{\emph{JHEP} {\bfseries 02}
  (2013) 039}, [\href{https://arxiv.org/abs/1210.1098}{{\ttfamily 1210.1098}}].

\bibitem{Nashed:uja}
G.~G.~L. Nashed, \emph{{A special exact spherically symmetric solution in f(T)
  gravity theories}},
  \href{https://doi.org/10.1007/s10714-013-1566-1}{\emph{Gen. Rel. Grav.}
  {\bfseries 45} (2013) 1887--1899},
  [\href{https://arxiv.org/abs/1502.05219}{{\ttfamily 1502.05219}}].

\bibitem{Stachel:1981fg}
J.~Stachel, \emph{{Globally stationary but locally static space-times: A
  gravitational analog of the Aharonov-Bohm effect}},
  \href{https://doi.org/10.1103/PhysRevD.26.1281}{\emph{Phys. Rev.} (1982)
  1281--1290}.

\bibitem{Wr}
R.~Weitzenb{\"o}k, \emph{Invarianten theorie}.
\newblock Noordhoff, Gr{\"o}ningen, 1923.

\bibitem{Nashed:2018cth}
G.~G.~L. Nashed and E.~N. Saridakis, \emph{{Rotating AdS black holes in
  Maxwell-$f(T)$ gravity}},
  \href{https://doi.org/10.1088/1361-6382/ab23d9}{\emph{Class. Quant. Grav.}
  {\bfseries 36} (2019) 135005},
  [\href{https://arxiv.org/abs/1811.03658}{{\ttfamily 1811.03658}}].

\bibitem{PhysRevD.15.2752}
G.~W. Gibbons and S.~W. Hawking, \emph{Action integrals and partition functions
  in quantum gravity},
  \href{https://doi.org/10.1103/PhysRevD.15.2752}{\emph{Phys. Rev. D}
  {\bfseries 15} (May, 1977) 2752--2756}.

\bibitem{Gibbons:1976pt}
G.~W. Gibbons and M.~J. Perry, \emph{{Black Holes and Thermal Green's
  Functions}}, \href{https://doi.org/10.1098/rspa.1978.0022}{\emph{Proc. Roy.
  Soc. Lond.} {\bfseries A358} (1978) 467--494}.

\bibitem{Gibbons:1978ac}
G.~W. Gibbons, S.~W. Hawking and M.~J. Perry, \emph{{Path Integrals and the
  Indefiniteness of the Gravitational Action}},
  \href{https://doi.org/10.1016/0550-3213(78)90161-X}{\emph{Nucl. Phys.}
  {\bfseries B138} (1978) 141--150}.

\bibitem{Nashed:2011fg}
G.~G.~L. Nashed, \emph{{Energy and momentum of a spherically symmetric dilaton
  frame as regularized by teleparallel gravity}},
  \href{https://doi.org/10.1002/andp.201100030}{\emph{Annalen Phys.} {\bfseries
  523} (2011) 450--458}, [\href{https://arxiv.org/abs/1105.0328}{{\ttfamily
  1105.0328}}].

\bibitem{Tipler:1977}
F.~J. Tipler, \emph{Singularities in conformally flat spacetimes},
  \href{https://doi.org/https://doi.org/10.1016/0375-9601(77)90508-4}{\emph{Physics
  Letters A} {\bfseries 64} (1977) 8 -- 10}.

\bibitem{Awad:2017sau}
A.~Awad and G.~Nashed, \emph{{Generalized teleparallel cosmology and initial
  singularity crossing}},
  \href{https://doi.org/10.1088/1475-7516/2017/02/046}{\emph{JCAP} {\bfseries
  1702} (2017) 046}, [\href{https://arxiv.org/abs/1701.06899}{{\ttfamily
  1701.06899}}].

\end{thebibliography}

\providecommand{\href}[2]{#2}\begingroup\raggedright\endgroup

\end{document}